\documentstyle[aps,prl]{revtex}

\begin{document}
\draft

\twocolumn[\hsize\textwidth\columnwidth\hsize\csname@twocolumnfalse\endcsname

\title{Quantum Breathers in Electron-phonon Systems}
\author{W. Z. Wang, A. R. Bishop, J. T. Gammel, and R. N. Silver}
\address{Theoretical Division and Center for Nonlinear Studies,
Los Alamos National Laboratory, Los Alamos, New Mexico 87545}

\date{\today}
\maketitle

\begin{abstract}
Quantum breathers are studied numerically in several electron-phonon
coupled finite chain systems, in which the coupling results in
intrinsic nonlinearity but with varying degrees of nonadiabaticity.  As
for quantum nonlinear lattice systems, we find that quantum breathers
can exist as eigenstates of the system Hamiltonians.  Optical responses
are calculated as signatures of these coherent nonlinear excitations.
We propose a new type of quantum nonlinear excitation,
breather-excitons, which are bound states of breathers and excitons,
whose properties are clarified with a minimal exciton-phonon model.
\end{abstract}

\ ]

\narrowtext

Much progress has been made toward understanding the physical
consequences of nonlinearity over the last decade.  In particular,
recent developments concerning ``breathers'' \cite{alan-rev} or
``intrinsic local modes'' (ILM) \cite{flach-prs}, suggest that energy
focusing is prevalent in both classical
\cite{flach-prs,page-book,aubry-prl,flach-3d} and quantum
\cite{jackiw-rmp,takeno-scott-aubry,wzw-qnll} {\it discrete} nonlinear
nonintegrable frameworks.  The existence and stability of multiquanta
bound states (breathers, ILMs) are now established for a wide variety
of discrete classical models with prescribed nonlinearity.  Recently,
we demonstrated that this property also persists in a discrete quantum
nonlinear lattice, and exhibited some distinctive observable signatures
in terms of spatial-temporal correlations \cite{wzw-qnll}.  However, of
more profound concern is the typical origin of effective nonlinearity
in quantum systems, namely, through the coupling of two or more
fields.  Adiabatic slaving of fields usually results in nonlinear
Schr\"odinger models.  Realistically, however, nonadiabatic effects
must be considered and the influences of nonlinearity and
nonadiabaticity are inevitably interrelated.  Here we consider examples
of electron-phonon ($e$-ph) coupled models frequently used to describe
organic and inorganic correlated electronic materials, for which an
adiabatic treatment of breathers is inadequate for various physical
observables \cite{horovitz-89}.  The interactions of electrons with the
lattice and among themselves provide sources of nonlinearity and
strongly influence electronic, optical and structural properties.
Here, we use numerical approaches restricted to finite chains but
without any adiabatic approximation.  We find that quantum breathers of
{\it two} types can exist in the $e$-ph systems considered --- those
near electronic ground states, and photoexcited breathers (termed
``breather-excitons'' below).

We consider a Holstein-Hubbard (HH) tight-binding model Hamiltonian
\cite{note-ssh} of an $e$-ph coupled system:
\begin{eqnarray}\label{eq1}
H_{e}\, &=&\, \displaystyle\sum_{i\sigma}\: -t_0\, (c_{i\sigma}^\dagger\,
c_{i+1,\sigma}\, +\, {\rm H.c.})\,\\
&& + \sum_i\: U\, n_{i\uparrow}\,n_{i\downarrow}\,
+ \sum_{i\sigma\sigma^\prime}\: V\, n_{i\sigma}\,n_{i+1\sigma^\prime} \nonumber\\
&&+\displaystyle\sum_i\: \hbar\omega_0\, \left( b_i^\dagger\, b_i\, +\,
{1\over 2} \right)
 \, -\, \displaystyle\sum_{i\sigma}\: \lambda\, 
\left( b_i^\dagger\, +\, b_i\right)\, n_{i\sigma}
\; .\nonumber
\end{eqnarray}
Here, $c_{i\sigma}^\dagger$ ($c_{i\sigma}$) and $b_i^\dagger$ ($b_i$)
are the creation (annihilation) operators of electrons and phonons,
respectively.  $t_0$, $U$ and $V$ are the electron kinetic energy, and
the on-site and nearest-neighbour Coulomb repulsions.  $\omega_0$ is
the bare phonon frequency.  The form of the $e$-ph coupling ($\lambda$
term) is that used in the Holstein (H) model \cite{holstein}.  We use
models with one electron per site ({\it i.e.} a ${1\over 2}$-filled
band).

\begin{figure}[]
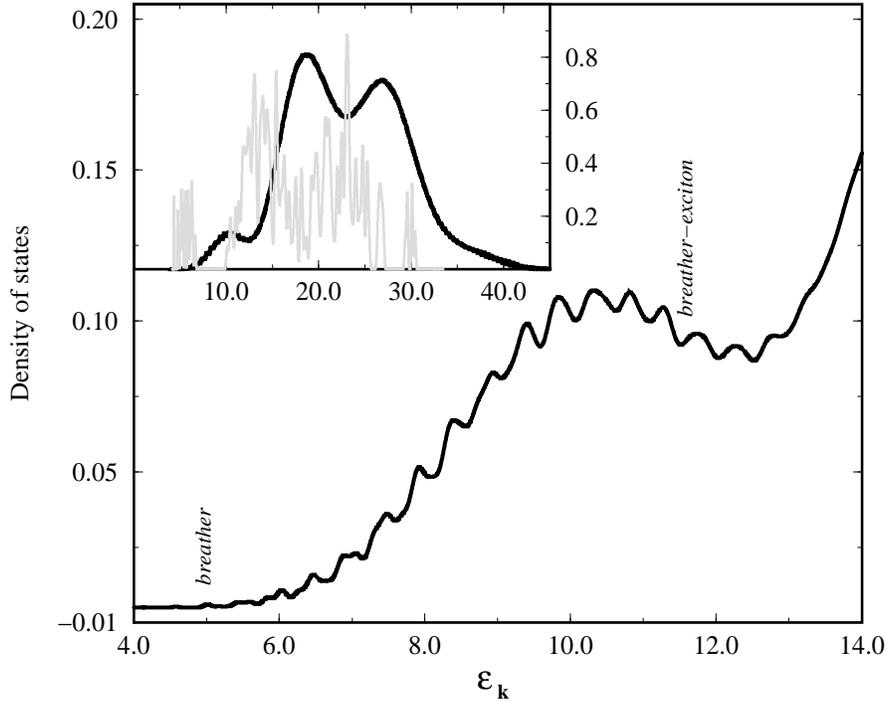

\caption[]{
Total DOS of a 6-site Holstein-Hubbard chain with $t_0=1.00,$
$\omega=0.40$, $\lambda=0.31$, $U=8.10$, $V=1.11$, $M_{\rm ph}=4$, and
${\cal M}=1\,638\,400$.  ($M_{\rm ph}$ and ${\cal M}$ stand for phonon
truncation and the total dimension of the matrix.) The inset shows the
DOS over the full energy range, with the corresponding pure Hubbard
model (gray line) results for comparison.
}\label{fig1}
\end{figure}

Our numerical approach mainly consists of exact diagonalization of
Hamiltonian matrices represented in Hilbert spaces defined by
appropriately selected basis functions, and the analysis of
characteristics of quantum breathers via various dynamic correlation
functions.  We deal fully with nonadiabaticity and $e$-$e$
correlations; the only approximation is the truncation of the infinite
phonon Hilbert space.  Because wavefunction information is needed to
identify a quantum breather state which is not necessarily low-lying in
the complete eigenspectrum, efficient exact diagonalization techniques
have been developed to handle large-scale matrices; these are also
designed to suit parallel computer architectures \cite{wzw-thesis}.  In
Fig.\ \ref{fig1}, we show a typical total density-of-states (DOS) of a
HH model, calculated using the kernel polynomial method \cite{rns-kpm}
which has a finite energy resolution.  The regions of interest are
indicated (see text below Figs.\ \ref{fig3} and \ref{fig5}).  Comparing
to the Hubbard model DOS, we realize that the denser the states the
more difficult they are to numerically distinguish, and the higher the
numerical efficiency required.  Also, since our model systems are
finite, it is not easy to eliminate finite-size effects
\cite{note-dmrg}.  However, our main results remain valid because the
breather excitations of interest here are intrinsically of finite
size.  There are also many realistic material contexts which are of
finite-size ({\it e.g.} conjugated oligomers).

\begin{figure}[]
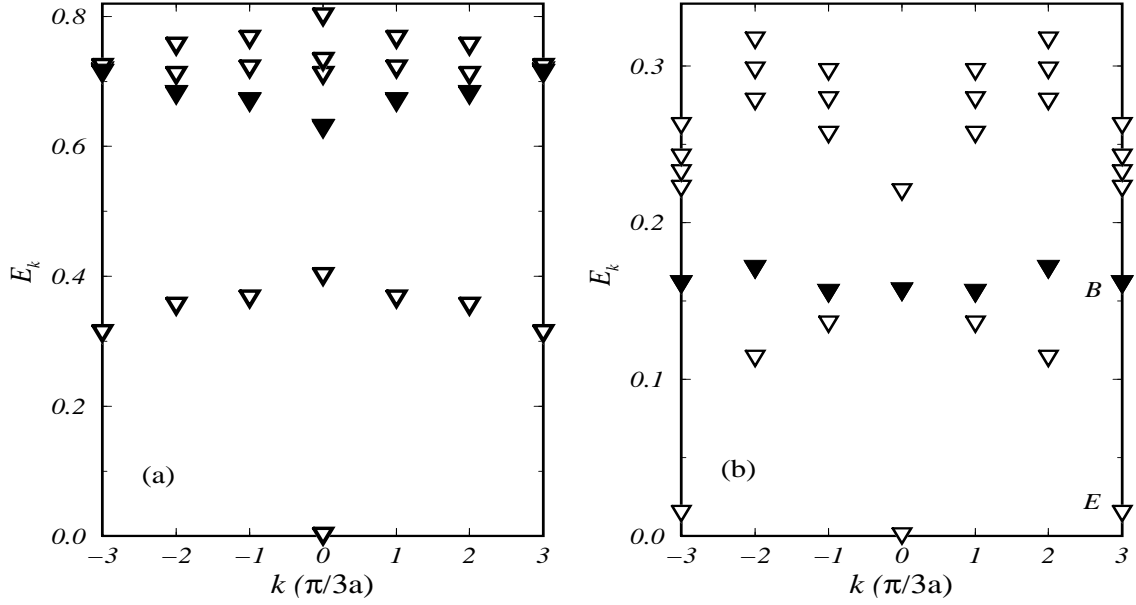

\caption[]{
Eigenspectrum (low-lying sector) of a 6-site Holstein chain with
$t_0$$=$$1.00$, $\omega$$=$$0.4$, $U$$=$$V$$=$$0.00$ and (a) $\lambda
=0.31$, $M_{\rm ph}(k=\pm{\pi\over 3},\,\pm{2\pi\over 3}) =5$, $M_{\rm
ph}(k=\pi)=6$, $M_{\rm ph}(k=0)=1$, ${\cal M}=1\,500\,000$; (b)
$\lambda =0.61$, $M_{\rm ph}(k=\pm{\pi\over 3},\,\pm{2\pi\over 3}) =6$,
$M_{\rm ph}(k=\pi)=10$, $M_{\rm ph}(k=0)=1$, ${\cal M}=5\,184\,000$.
The breather states are labeled with solid triangles, and the remaining
ones with open triangles.  The ground state energy has been subtracted
from $E_k$.
}\label{fig2}
\end{figure}

\begin{figure}[]
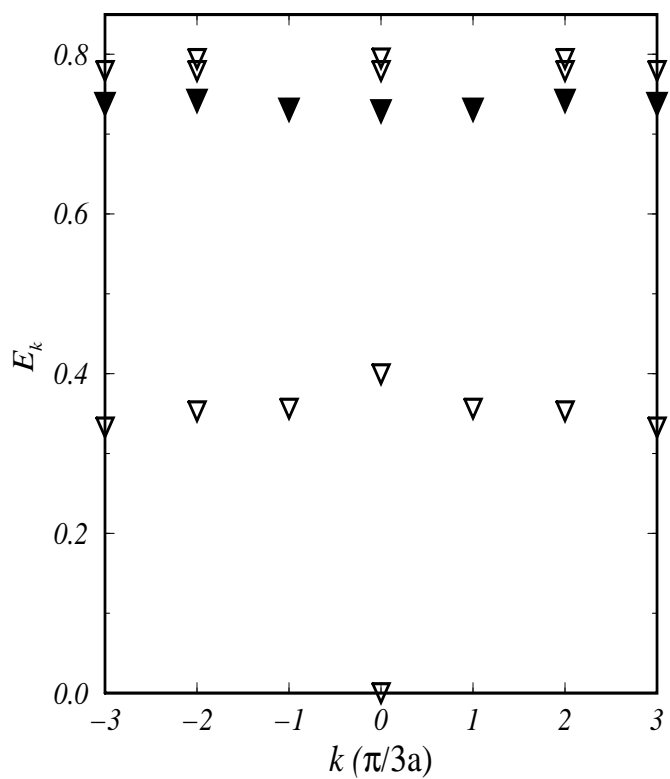

\caption[]{
Eigenspectrum of the 6-site Holstein-Hubbard chain in
Fig.\ \ref{fig1}.  The notation is as in Fig.\ \ref{fig2}.  The ground
state energy is 4.14$t_0$
}\label{fig3}
\end{figure}

\begin{figure}[h]
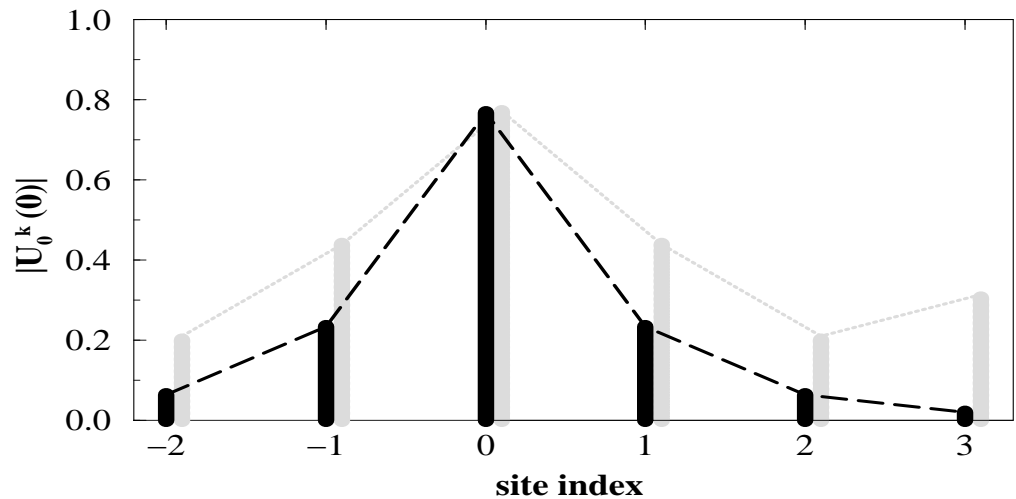

\caption[]{
The spatial correlation function $U_j^k(0)$ of the system in
Fig.\ \ref{fig2}(b).  The solid line is for a breather state (labeled
{\it B} in Fig.\ \ref{fig2}(b)), while the gray line is for an extended
state ({\it E}).
}\label{fig4}
\end{figure}

\begin{figure}
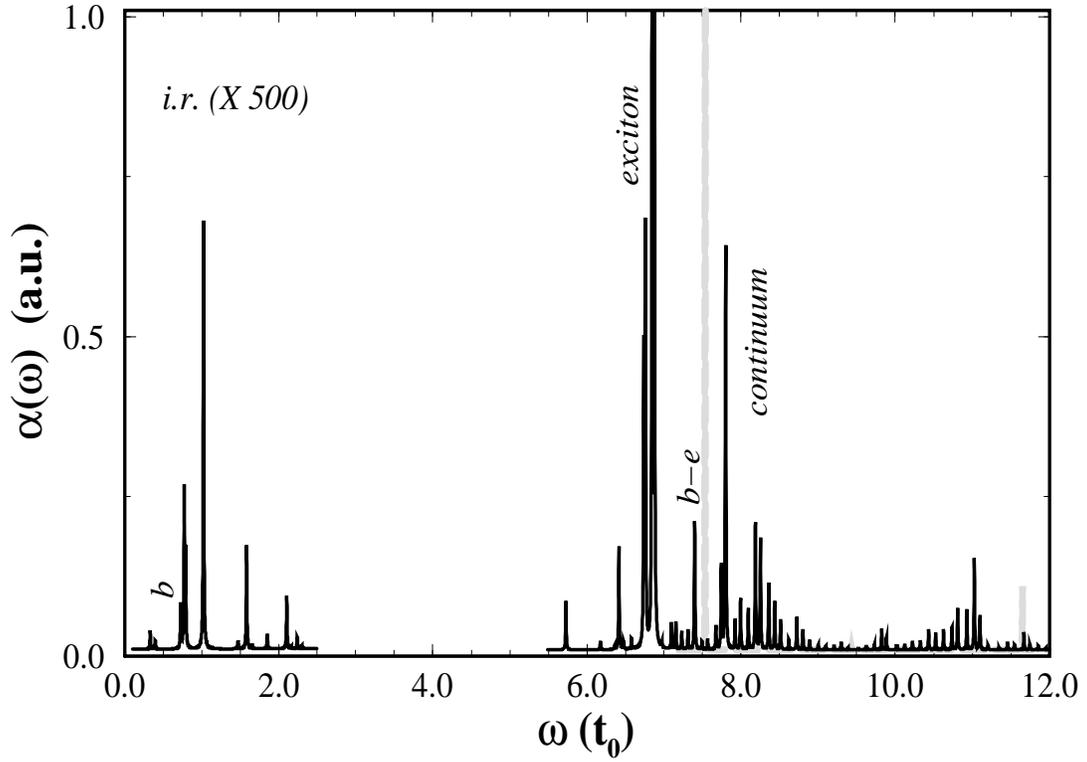

\caption[]{
The zero-temperature infrared and electronic optical absorptions of the
6-site Holstein-Hubbard system in Fig.\ \ref{fig3} (solid line) and the
related 6-site Hubbard model with $t_0=1.00$, $U=8.10$, $V=1.11$ (gray
line).  All spectra are broadened by a Lorenzian of width 0.005.
\label{fig5}}
\end{figure}

The eigen-energies of the system (\ref{eq1}) are shown in
Figs.\ \ref{fig2} and \ref{fig3} for 6-site Holstein, and HH models,
respectively.  The results in Fig.\ \ref{fig2} were obtained using a
basis set of Debye phonons and Bloch electron functions, while
Fig.\ \ref{fig3} results use Einstein phonons and Wannier functions.
The degrees of softening of the $k=\pi$ mode (Kohn anomaly) indicate
that the systems in Figs. 2(a) and 3 are weakly $e$-ph coupled, whereas
that in Fig.\ \ref{fig2}(b) is more strongly coupled.  The phonon
dispersion is caused by effective nonlinear ph-ph coupling via the
$e$-ph coupling.  Because of discrete translational invariance, direct
examination of the wavefunction amplitudes and single phonon operator
expectation values do not reveal the presence of quantum breathers
\cite{wzw-qnll}.  Rather, we examine various static and dynamic
correlation functions of lattice displacements, including:
$U_j^k(t)$$=$$\sum_i$$\left\langle k\left|\right.\right.$
$u_i(0)$$u_j(t)$ $\left.\left.\right| k \right\rangle$, where $u_i$ is
the phonon displacement operator.  The static correlation functions
($t$=$0$) probe the spatial localization in a given eigenstate, while
the dynamic counterparts probe the temporal coherence.  These two
properties are the distinguishing characteristics of quantum breathers
\cite{alan-rev,flach-prs}.  The correlation functions $U_j^k(t)$, as in
Fig.\ \ref{fig4}, show that there indeed exists a band of particle-like
states in each of the above examples, and that these states possess
short but finite spatial and temporal correlations, whereas all other
states are extended \cite{wzw-rev}.  A typical case is shown in
Fig.\ \ref{fig2}(b), where the particle-like band possesses a large
binding energy separating it from the continuum bands above, and the
anharmonicity and corresponding localization are strong (the
correlation length is approximately 3 lattice constants, see
Fig.\ \ref{fig4}) \cite{note-ssh}.

The existence and properties of quantum breathers depend not only on
the effective nonlinear ph-ph coupling (as in \cite{wzw-qnll}) but also
on the nonadiabaticity (here, controlled by the ratio $\hbar\omega_0/
t_0$).  We find that the stronger the $e$-ph coupling and the
adiabaticity, the more easily quantum breathers form\cite{wzw-rev}.
Furthermore, these breather states can survive strong $e$-$e$
correlations (Fig.\ \ref{fig3}), although we find that in the lowest
part of the excitation spectrum strong $e$-$e$ correlations tend to
induce extended magnetic excitations \cite{wzw-rev}.  We will see
below, however, that strong $e$-$e$ correlations do not necessarily
destroy short-correlation length breathers in other spectral regions.


To identify physical consequences of the breather excitations, we study
the first-order optical response function which can be measured in
one-photon experiments.  Fig.\ \ref{fig5} shows the zero-temperature
infrared and electronic optical absorptions of a HH system.  Within
this nonadiabatic approach, we account for all the electronic and
phonon polarizabilities within one photon perturbations.  This is a
step toward understanding accumulating experimental results in
ultra-fast time-resolved (nonadiabatic) and nonequilibrium measurements
\cite{valy-97}.  First, we observe that in the infrared region, several
prominent peaks directly indicate the anharmonicity, including
multiphonon side-bands.  Comparing to Fig.\ \ref{fig3}, the
contribution to the infrared absorption from the breather states
(marked ``{\it b}'') has comparable intensity to that from surrounding
extended states.  Second, below the main absorption peak (marked ``{\it
exciton}''), there exist a series of spectral features which cannot be
explained in an adiabatic description.  To understand their origin, we
added small amounts of disorder in the electron or phonon degrees of
freedom ({\it e.g.}, the parameters $U$ or $\hbar\omega_0$).  We found
that only phonon disorder changes the spectral features near the ``{\it
exciton}'' edge \cite{wzw-rev}, indicating their phonon origin.  Third,
in addition to the rich structure in the region of electronic
absorption, we observe a series of bands between the ``{\it exciton}''
edge and the ``{\it continuum}'' bands.  Among them, we notice one band
(``{\it b-e}'') whose position and intensity relative to the other
phonon sidebands does not change under the perturbation of phonon
disorder \cite{wzw-rev}.  This indicates that this particular band is
due to some a relatively stable excited configurations which are more
localized.  This evidence suggests a possible new type of bound state %
which is higher in energy than the lowest excitons and more localized
than the nearby states --- we will term these {\it breather-excitons}.
Below we show that they possess some of the characteristics of the
``ground state'' quantum breathers identified in the low-lying part of
the eigenspectrum [Fig.\ \ref{fig3}].

Physical intuition suggests the possibility of a bound state of
excitons and breathers, {\it i.e.} a hot (dressed) exciton or a
photoexcited breather.  Excitons exist primarily as electron-hole pairs
bound by both the Coulomb interactions and $e$-ph coupling.  The latter
factor slows down the exciton motion and tends to dress the exciton
with phonons and breathers.  $e$-$e$ correlations provide another
energy region from which the breather can be excited, and contribute
additional nonlinearity enhancing the breather formation.  Furthermore,
with the excited oscillating dipole moments inside the breather-exciton
bound states, they will strongly absorb photons, as an electronic
exciton does.

\begin{figure}[h]
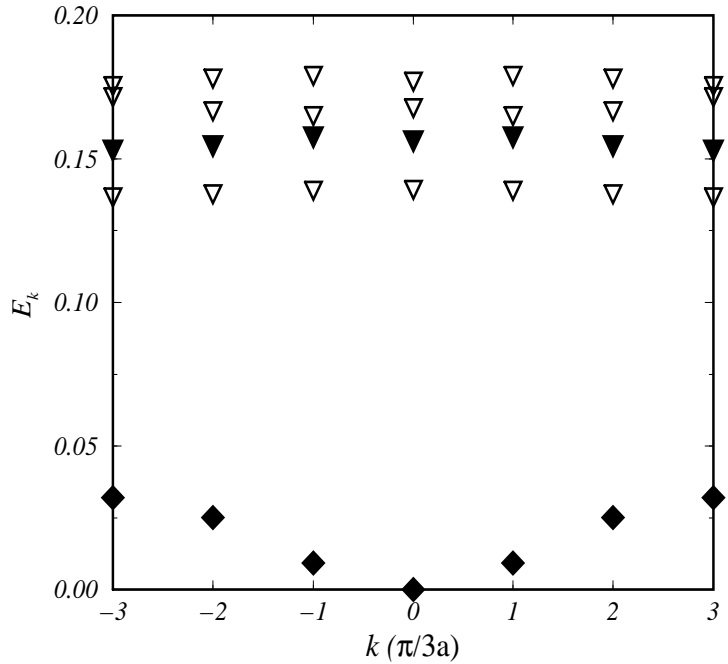

\caption[]{
Eigenspectrum (low-lying sector) of the 6-site minimal exciton-phonon
model (see text) with $t_e=1.00$, $t_h=0.67$, $\epsilon_e=4.00$,
$\epsilon_h=-4.00$, $\omega_0 =0.15$, $U=18.0$, $V=2.75$, $\lambda_e =
0.87$, $\epsilon_h=-0.27$.  $M_{\rm ph}=6$, ${\cal M}=1\,679\,616$.
The breather-exciton states are labeled with solid triangles, the
dressed excitons with diamonds, and the remaining ones with open
triangles.  The ground state energy has been subtracted from $E_k$.
}\label{fig6}
\end{figure}

To support the above arguments \cite{note-high}, we introduce a minimal
exciton-phonon model, describing an electron and hole interacting with
each other and with phonons:
\begin{eqnarray}\label{eq2}
H\, &=&\, \displaystyle\sum_i\: -\, t_e\, \left(\,
e_i^\dagger\, e_{i+1}\, +\, {\rm H.c.}\,\right)\: +\, \sum_{i}\: \epsilon_e\,
e_{i}^\dagger\, e_i\; \nonumber\\
&&\displaystyle\sum_i\: -\, t_h\, \left(\,
h_i^\dagger\, h_{i+1}\, +\, {\rm H.c.}\,\right)\: -\, \sum_{i}\: \epsilon_h\,
h_{i}^\dagger\, h_i\; \nonumber\\
&& + \displaystyle\sum_{i,j}\: V(i-j)\,
e_i^\dagger e_i \, h_j^\dagger h_j\nonumber\\
&& - \sum_{i}\: \lambda_e\, \left( b_i^\dagger\, + b_i\right)\, 
e_i^\dagger\, e_i
- \sum_{i}\: \lambda_h\, \, \left( b_i^\dagger\, + b_i\right)\,
h_i^\dagger\, h_i\nonumber\\
&&+\displaystyle\sum_i\: \hbar\omega_0\, \left( b_i^\dagger\, b_i\, +\,
{1\over 2} \right) \: ,
\end{eqnarray}
with $V(i$$-$$j)$ $=$ $-U$ (as $i$ $=$ $j$); $-V$ (as $|i$$-$$j|$ $=$
$1$); and $0$ (otherwise).  $e_i^\dagger$ ($e_i$) and $h_i^\dagger$
($h_i$) are creation (annihilation) operators for the electron and
hole, respectively.  $t_{e(h)}$, $\epsilon_{e(h)}$, $V(i-j)$ and
$\lambda_{e(h)}$ are the electron (hole) hopping integrals, on-site
energies, electron-hole attractions and $e(h)$-ph couplings strengths,
respectively \cite{note-brex}.  Fig.\ \ref{fig6} shows the low-lying
part of the eigenspectrum, corresponding to the ``{\it exciton}''
region in Fig.\ \ref{fig5}.  The lowest band corresponds to the lowest
exciton states in Fig.\ \ref{fig5} and shows dispersion.  They are
excitons dressed by a static lattice distortion of condensed phonons.
The breather-exciton states are also strongly localized, but dressed
with a dynamic phonon wavepacket (weak dispersion) and exhibiting more
coherent internal motions. (We speculate that the excitonic internal
frequency resonantly traps a breather.) The remaining states are
$e$-$h$ pairs coupled to extended anharmonic phonons.
Breather-excitons are more polarized than the ground state breathers,
leading to the stronger optical absorption seen in Fig.\ \ref{fig5}.

In summary, this study is an important extension of previous
investigation of quantum breathers in nonlinear lattices
\cite{wzw-qnll}.  It demonstrates their existence in more general
$e$-ph systems where nonlinearity is self-consistently generated by the
$e$-ph (and $e$-$e$) coupling rather than being inserted by hand.  Thus
the combined influences of nonadiabaticity and nonlinearity are
incorporated.  Clear features from breathers are found in optical
properties.  The concept of {\it breather-exciton} states was proposed
and demonstrated for a simplified exciton-phonon model without any
adiabatic approximations.  Investigations are underway \cite{wzw-abg}
to correlate our findings with nonlinear optics and time-resolved
spectroscopy, including the competing timescales associated with
exciton formation, excitonic self-trapping, and breather-exciton
formation, as functions of nonlinearity and nonadiabaticity
\cite{wzw-rev}.

We are grateful to M. I. Salkola and J. Zang for many useful
discussions and previous collaborations.  We also thank S. R White, C.
L. Zhang, Z. V. Vardeny, L. Yu and Z. B. Su for helpful discussions.
This computation was performed at the CM-5 of ACL and the Sun UE-4000
of T-11 at Los Alamos National Laboratory.  The research is supported
by the U.S. Department of Energy.

\begin{references}
\bibitem{note-ssh} We have obtained similar results for the SSH model \cite{ssh,wzw-rev}.
\bibitem{note-brex} Only simplified $e(h)$-ph and $e$-$h$ interactions
are retained in Eq. (2) because no exact mapping is available between
an effective exciton-phonon model and the original half-filled HH model.
\end{references}
\end{document}